\documentclass[fleqn,usenatbib]{mnras}

\usepackage{newtxtext,newtxmath}

\usepackage[T1]{fontenc}
\usepackage{ae,aecompl}

\usepackage{graphicx}	
\usepackage{amsmath}	
\usepackage{amssymb}	
\usepackage{hyperref}

\title[Feeding and Feedback in AGN]{Gemini NIFS survey of feeding and feedback in nearby Active Galaxies - III. Ionized versus warm molecular gas masses and distributions}

\author[Astor J. Sch\"onell Jr.]{Astor J. Sch\"onell Jr.$^{1,3}$\thanks{E-mail:
juniorfisicoo@gmail.com}, Thaisa Storchi-Bergmann$^{1}$, Rogemar A. Riffel$^{2}$, 
\newauthor Rog\'erio Riffel$^{1}$,  Marina Bianchin$^2$, Luis G. Dahmer-Hahn$^{1}$, Marlon R. Diniz$^{2}$,
 \newauthor  Natacha Z. Dametto$^{1}$ \\
$^{1}$Instituto de F\'isica, Universidade Federal do Rio Grande do Sul, Av. Bento Gon\c calves 9500, 91501-970\\ 
Porto Alegre, RS, Brazil \\
$^{2}$Universidade Federal de Santa Maria, Departamento de F\'isica, Centro de Ci\^encias Naturais e Exatas, 97105-900,\\
Santa Maria, RS, Brazil\\
$^3$ Instituto Federal de Educa\c c\~ao, Ci\^encia e Tecnologia Farroupilha, BR287, km 360, Estrada do Chapad\~ao, 97760-000, \\
Jaguari - RS, Brazil
}
\date{Accepted XXX. Received YYY; in original form ZZZ}

\pubyear{2019}

\begin{document}
\label{firstpage}
\pagerange{\pageref{firstpage}--\pageref{lastpage}}
\maketitle

\begin{abstract}

We have used the Gemini Near-Infrared Integral Field Spectrograph (NIFS) in the J and K bands to map the distribution, excitation and kinematics of the ionized H\,{\sc ii} and warm molecular gas H$_2$, in the inner few 100\,pc of 6 nearby active galaxies: NGC\,788, Mrk\,607, NGC\,3227, NGC\,3516, NGC\,5506, NGC\,5899. {For most galaxies, this is the first time that such maps have been obtained}. The ionized and H$_2$ gas show distinct kinematics: while the H$_2$ gas is mostly rotating in the galaxy plane with low velocity dispersion ($\sigma$), the ionized gas usually shows signatures of outflows associated with higher $\sigma$ values, most clearly seen in the [Fe\,{\sc ii}] emission line. These two gas species also present distinct flux distributions: the H$_2$ is more uniformly spread over the whole galaxy plane, while the ionized gas is more concentrated around the nucleus and/or collimated along the ionization axis of its Active Galactic Nucleus (AGN), presenting a steeper gradient in the average surface mass density profile than the H$_2$ gas. The total H\,{\sc ii} masses cover the range $2\times\,10^5-2\times\,10^7$\,M$_{\odot}$, with surface mass densities in the range 3--150\,M$_{\odot}$\,pc$^{-2}$, while for the warm H$_2$ the values are 10$^{3-4}$ times lower. We estimate that the available gas reservoir is at least $\approx$\,100 times more massive than needed to power the AGN. If this gas form new stars the star-formation rates, obtained from the Kennicutt-schmidt scalling relation, are in the range 1--260$\times$ 10$^{-3}$ M$_{\odot}$ yr$^{-1}$. But the gas will also -- at least in part -- be ejected in the form of the observed otflows.



\end{abstract}




\section{Introduction}

The growth of super-massive black holes (SMBH) and their host galaxies are connected by the AGN feeding and feedback processes, that can presumably explain the correlation between the mass of the SMBH and the mass of the galaxy bulge \citep{ferrarese2005,somerville2008,kormendy2013}. The feeding via gas accretion is required to trigger the nuclear activity, while the feedback provided by the AGN radiation and outflows is fundamental to constrain galaxy evolution models, since without the AGN feedback the models predict that the most massive galaxies form too many stars and grow more than observed \citep{springel2005,fabian2012,terrazas2016}).

Gas distribution, excitation and kinematics in the vicinity of Active Galactic Nuclei (AGN) ($\approx$\,100 pc scales) provide important constraints on the physics of the AGN feeding and feedback processes. The near-infrared (hereafter, near-IR) integral field spectroscopy (IFS) of nearby galaxies is an effective method to quantify these processes. {Eight to ten-meter telescopes IFS with adaptive optics (AO) can provide two-dimensional coverage with spatial resolution of a few to tens of parsecs in nearby galaxies at spectral resolutions  that allow to resolve gas inflows and outflows at such scales \citep{muller09,davies09,davies2014}.}
{AO systems are available mainly in the near-IR, a spectral region that also has the advantage of being less affected by dust extinction (usually high in the central region of galaxies) than optical observations. The use of AO assisted IFS of nearby galaxies in the near-IR thus allows to resolve the central regions down to a few parsecs, and also simultaneously map two distinct gas phases: the ionized and molecular (H$_2$) gas emission. The latter is not available in the optical but can be observed in the near-IR K band.}

The near-IR line-emission at $\approx$\,100\,pc scales in AGN hosts is originated by the heating and ionization of ambient gas by the AGN radiation and by shocks produced by radio jets \citep{riffel06a,rsn10}. Recent observations by our group AGNIFS - AGN Integral Field Spectroscopy \citep[e.g.][]{riffel2018} and others show that the {molecular and ionized} gas have distinct spacial distributions and kinematics at these scales: {the former is usually more restricted to the plane of galaxies, with the kinematics being dominated by rotation in the disk in most cases, and presenting also signatures of inflows in some cases}; the latter traces a more disturbed medium, usually associated to outflows from the AGN, but frequently showing also a disk rotation component \citep[e.g.][]{rsn10,riffel2013,mazzalay2014,barbosa14,diniz2015}. Our previous studies led to the conclusion that, while the ionized gas emission can be considered a tracer of the AGN feedback, the molecular gas emission is usually a tracer of its feeding.

 In this paper we present maps of the ionized and molecular gas distribution, excitation and kinematics of the inner 3$^{\prime\prime}\times$3$^{\prime\prime}$ of a sample of 6 nearby Seyfert galaxies. The discussion is restricted to the the gas mass distributions and total ionized and molecular gas masses as well as to the presentation of the global gas kinematics, pointing out signatures of rotation and outflows. The analysis and discussion of the gas excitation as well as the modelling of the gas kinematics and quantification of outflows will be deferred to a forthcoming paper (hereafter identified as Paper B). This work is the third paper with the results of a large Gemini proposal (P.I. Storchi-Bergmann) in which our group AGNIFS aims to map and quantify the feeding and feedback processes of a sample of 29 nearby Seyfert galaxies \citep{riffel2018}, 
selected for their proximity and X-ray luminosity, as described {in Sect. \ref{sec:sample}}. Our ultimate goal is to investigate possible correlations between measured properties (as gas masses and densities, mass inflow and outflow rates and kinetic power of the outflows) and the AGN luminosity. Results for individual galaxies of the sample have been already  presented in previous papers by our group: NGC\,4051 \citep{riffel08,riffel2017}, NGC\,4151 \citep{storchi2009,storchi10,riffel09}, Mrk\,1066 \citep{riffel10,rs1157,ramos2009,riffel2017}, Mrk\,1157 \citep{rs1157,riffel2011b,riffel2017}, NGC\,1068 \citep{storchi2012,riffel2014,barbosa14}, 
Mrk\,79 \citep{riffel2013}, Mrk\,766 \citep{astor2014,riffel2017}, NGC5929 \citep{riffel5929,riffel2015,riffel2017}, NGC\,2110 \citep{diniz2015}, NGC\,5548 \citep{astor2017,riffel2017}, 
NGC\,788, NGC\,3227, NGC\,3516, NGC\,4235, NGC\,4388, NGC\,5506, NGC\,1052, NGC\,5899 and Mrk\,607 \citep{riffel2017}. 

This paper is organized as follows. In Section\,\ref{sec:sample} we present the sample {and in Section\,\ref{sec:obs} the} description of the observations and data reduction procedures, while the fitting procedure of the emission lines is discussed in Section\,\ref{sec:measure}. The results are shown in Section\,\ref{sec:results} , we discuss them in Section\,\ref{sec:discussion} , and in Section\,\ref{sec:conclusions} we present our conclusions.

\section{Sample}
\label{sec:sample}
Our {AGN sample} was selected from the Swift-BAT 60-month catalogue adopting three criteria: (I) 14$-$195 keV luminosities $L_{X}\ge$ 10$^{41.5}$ erg s$^{-1}$, (II) redshift $z\le0.015$, 
and (III) being accessible for NIFS ($-$30$^{o}$ $<$ $\delta$ $<$ 73$^{o}$). The selectoin according to the hard (14$-$195 keV) band emission of the Swift-BAT survey is justified by the fact that it  measures direct emission from the AGN rather than scattered or re-processed emission, and is much less sensitive to obscuration in the line-of-sight than soft X-rays or optical observations, allowing a selection based only on the AGN properties. In order to assure that we will be able to probe the feeding and feedback processes we further selected the galaxies for having previously observed extended [O\,{\sc iii}]$\lambda$5007 emission \citep{schmitt2000}, which enhances the probability of them presenting extended near-IR line emission, needed to map the gas excitation and kinematics. We have excluded a few galaxies that had guiding problems in the observations and included additional targets from our previous NIFS observations to complement the sample{, leading to a total sample of 29 active galaxies}. A complete characterization of the sample has been presented in \citet{riffel2018}. The observations are still in progress and will probably be concluded in 2019. So far, 20 galaxies have been observed and  here we show the gas distribution and excitation for 6 of them. {These six galaxies} are listed in Table\,\ref{tabsample}, which presents also their nuclear activity type, morphological classification and information about the observations. 

\section{Observations and data reduction}
\label{sec:obs}

We used J and K band data obtained with the Gemini Near-Infrared Integral-Field Spectrograph (NIFS; \citet{mcgregor2003}) with the AO system ALTAIR between 2008 and 2016. NIFS has a square field of view of 3$^{\prime\prime}\times$3$^{\prime\prime}$, divided
into 29 slitlets 0\farcs103 wide with a spatial sampling of
0\farcs042 along them.  We used the standard Sky--Object--Object--Sky dither sequence in the observations, with off-source sky positions since all targets are extended. The individual exposure times varied according to the target and are listed, together with further information on the observations in Table\,\ref{tabsample}. The filter ZJ\_G0601 was used with the J-band observations, while the K-band observations were performed using the HK\_G0603 filter, {as shown in Table\,\,\ref{tabobs} together with further details of the instrument configuration}.

\begin{table*}
\centering

\vspace{0.3cm}
\caption{Log of the observations together with basic information on the sample galaxies. (1) Galaxy name; (2) Project ID; (3) J and (4) K-band exposure times (s); (5) Distance (Mpc); (6) Nuclear Activity; (7) Hubble type from NED;  (8) Scale; (9) AGN bolometric luminosity.}
\begin{tabular}{c c c c c c c c c}
\hline
1 & 2 & 3 & 4 & 5 & 6 & 7&8 & 9\\
Galaxy   & Project ID & J Exp-time & K Exp-time  & D   & Act. & Hub. type &  Scale & logL$_{\rm AGN}$\\
	&	& (sec)  &  (sec) & (Mpc)  &  & &  (pc/arcsec) & (erg/s)\\
\hline
NGC788  & GN-2015B-Q-29 & $7\times400$ & $11\times400$ & 56.1 & Sy2 & SA0/a(s)   & 272 & 44.4 \\
Mrk607  & GN-2012B-Q-45& $10\times500$  & $12\times500$ & 36.1 & Sy2 & Sa edge-on & 175 & - \\
NGC3227 & GN-2016A-Q-6 & $6\times400$ & $6\times400$    & 14.8 & Sy1.5 & SAB(s)a pec &  72 & 43.4 \\
NGC3516 & GN-2015A-Q-3 & $10\times450$ & $10\times450$ & 38.0 & Sy1.5 & (R)SB0$^0$(s)  & 184 & 44.2 \\
NGC5506 & GN-2015A-Q-3 & $10\times400$ & $10\times400$ & 24.9 & Sy1.9 & Sa pec edge-on &  121 & 44.3 \\
NGC5899 & GN-2013A-Q-48& $10\times460$& $10\times460$& 36.8 & Sy2 & SAB(rs)c &  178 & 43.1\\
\hline
\end{tabular}
\label{tabsample}
\end{table*}
  
\begin{figure*}
\includegraphics[width=0.98\textwidth]{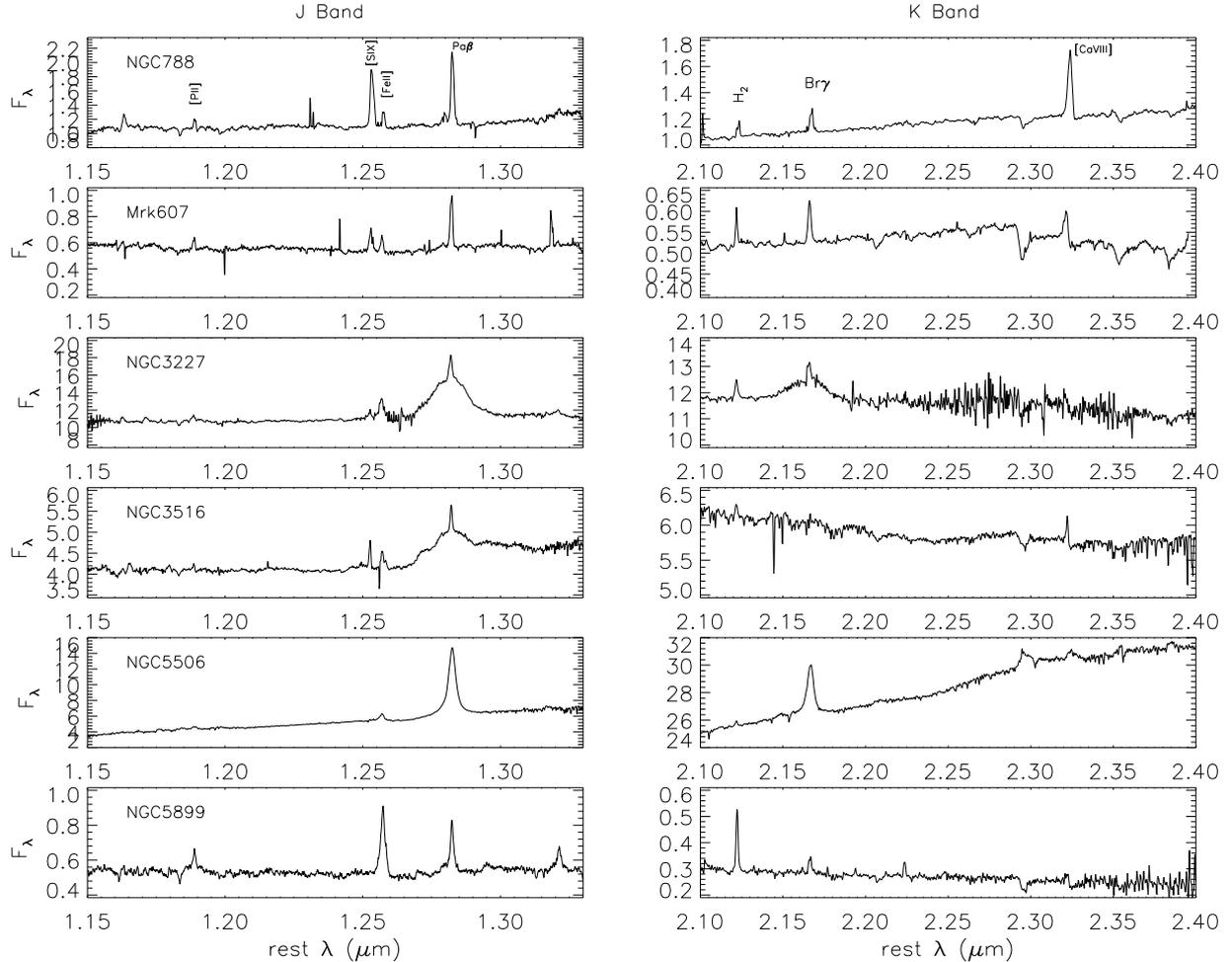}
\caption{J-band (left) and K-band (right) pectra obtained within a 0\farcs25$\times$0\farcs25 aperture centred at the nucleus of the galaxy indicated in the top left corner of the left panels. Flux units are 10$^{-16}$\,erg\,s$^{-1}$\,cm$^{-2}$\,\AA$^{-1}$.}
\label{espectros}
\end{figure*}

Data reduction followed standard procedures and was accomplished using tasks specifically developed for NIFS data reduction, as part of \texttt{gemini IRAF} package, as well as generic \texttt{IRAF} tasks and IDL scripts. The procedures included trimming of the images, flat-fielding, sky subtraction, wavelength and s-distortion calibrations. The telluric absorptions have been removed using A-type standard stars observations. These stars were also used to flux calibrate the spectra of the galaxies by interpolating a black body function to the spectrum of each star in order to generate the sensitivity function. Finally, calibrated datacubes were created for each individual exposure at an angular sampling of 0\farcs05$\times$0\farcs05 and combined in a final datacube for each galaxy. All datacubes cover the inner $\approx$ 3\farcs0$\times$3\farcs0 and have typical spectra as shown in Fig.\,\ref{espectros}. As shown in \citet{riffel2018}, the spectral resolutions for both bands correspond to velocities of about 40\,km\,s$^{-1}$ and the angular resolution ranges from 0\farcs12 to 0\farcs18, corresponding to a few tens of parsecs at the galaxies.

\begin{table*}
\vspace{0.3cm}
\caption{Configuration of the observations: (1) spectral band, (2) grating, (3) filter, (4) filter central wavelength, (5) filter spectral range, (6) galaxies observed in each configuration. }
\begin{tabular}{l c c c c c}
\hline
Band  & Grating & Filter & Central Wav. & Spec. Range  & Galaxies observed \\
  &  &  & ($\mu$m) &  ($\mu$m) &  \\
\hline
J	& J\_G5603  & ZJ\_G0601 &  1.25   &   1.14 -- 1.36   & All		\\
K	& K\_G5605  & HK\_G0603 &  2.20   &    1.98 -- 2.40   & NGC\,788, NGC\,3516 and NGC5506 \\
K$_l$ & Kl\_G5607 & HK\_G0603 &2.30   &  2.08 -- 2.50   &  Mrk\,607, NGC\,3227 and NGC\,5899\\
\hline
\end{tabular}
\label{tabobs}
\end{table*}



\section{Measurements}
\label{sec:measure}
We have used the {\sc PROFIT} routine \citep{profit} to fit the profiles of the following emission lines at each pixel over the whole field-of-view (FOV): [P\,{\sc ii}]$\lambda$1.1886$\mu$m, [Fe\,{\sc ii}]$\lambda$1.2570$\mu$m, Pa${\beta}\lambda$1.2822$\mu$m, H$_2 \lambda$2.1218$\mu$m and Br${\gamma} \lambda$2.1661$\mu$m. This was done using Gauss-Hermite series, which were chosen to preserve most of the gas velocity information by fitting also the emission line wings via the moments $h_3$ and $h_4$, besides returning the line-of-sight velocity (V$_{\rm LOS}$) and velocity dispersion ($\sigma$).

The $h_3$ Gauss-Hermite moment measures asymmetric deviations from a Gaussian profile, such as blue (negative values) or red (positive 
values) wings, while the $h_4$ moment quantifies the peakness of the profile, with positive values for a more peaked profile and more extended wings than a Gaussian and negative values for a broader profile (more flat-topped) and with less extended wings  than that of a Gaussian curve.

The underlying continuum was fitted using a first degree function, since the spectral range used in the fit of each emission line was small. The routine uses the  {\sc mpfitfun} routine \citep{mark09} to perform a non-linear $\chi^2$ minimization. The fit of the line plus continuum involves 7 free parameters -- line amplitude, central wavelength, $\sigma$, $h_3$, $h_4$ plus another 2 from the first degree function used to fit the continuum. The only restrictions placed are that $|h_3|$ and $|h_4|$ are $<$ 0.5, but most returned values smaller than these. Actually, this restriction was hardly used, as values 0.5 or higher indicate profiles that are very distinct from Gaussian curves, which are not observed in the galaxies' spectra. The {\sc profit} routine also outputs 1-sigma errors for each of the parameters, computed from the covariance matrix.

In the case of NGC\,3227, NGC\,3516 and NGC\,5506 we also fitted a broad component to the Pa$\beta$ and Br$\gamma$ emission line profiles in the central region, as these galaxies host type 1 AGN. This was done through a modification of the {\sc PROFIT} routine to fit the broad component and subtract its contribution from the profiles in order to generate a data cube only with the narrow components. The steps for such task were: ({\sc i}) fit of only one Gaussian to the broad component, by masking strong narrow emission lines; ({\sc ii}) its subtraction from the spectra in which it is present, and ({\sc iii}) fit of the narrow components. We achieved very satisfactory fits in all cases, with no constraints placed for the fit. The spatial region within which the fit of a broad component was necessary is shown by a cyan square in the third panel of the first row in Figs. \,\ref{ngc3227}, \ref{ngc3516} and \ref{ngc5506}, respectively for NGC\,3227, NGC\,3516 and NGC\,5506. 
As the broad components are from the unresolved Broad Line Region (BLR), during the fit we kept fixed the central wavelength and width of the broad components to the value measured from the integrated spectrum within the square mentioned above and allowed only the variation of its amplitude. From the measurements presented in Figs. \,\ref{ngc3227}, \ref{ngc3516} and \ref{ngc5506}, we conclude that the subtraction of the broad components was satisfactory, as we found no traces of them in the subtracted data cube.

\section{Results}
\label{sec:results}
\begin{figure*}
\begin{tabular}{c}
\bf{\LARGE NGC\,788} \\

\includegraphics[width=0.98\textwidth]{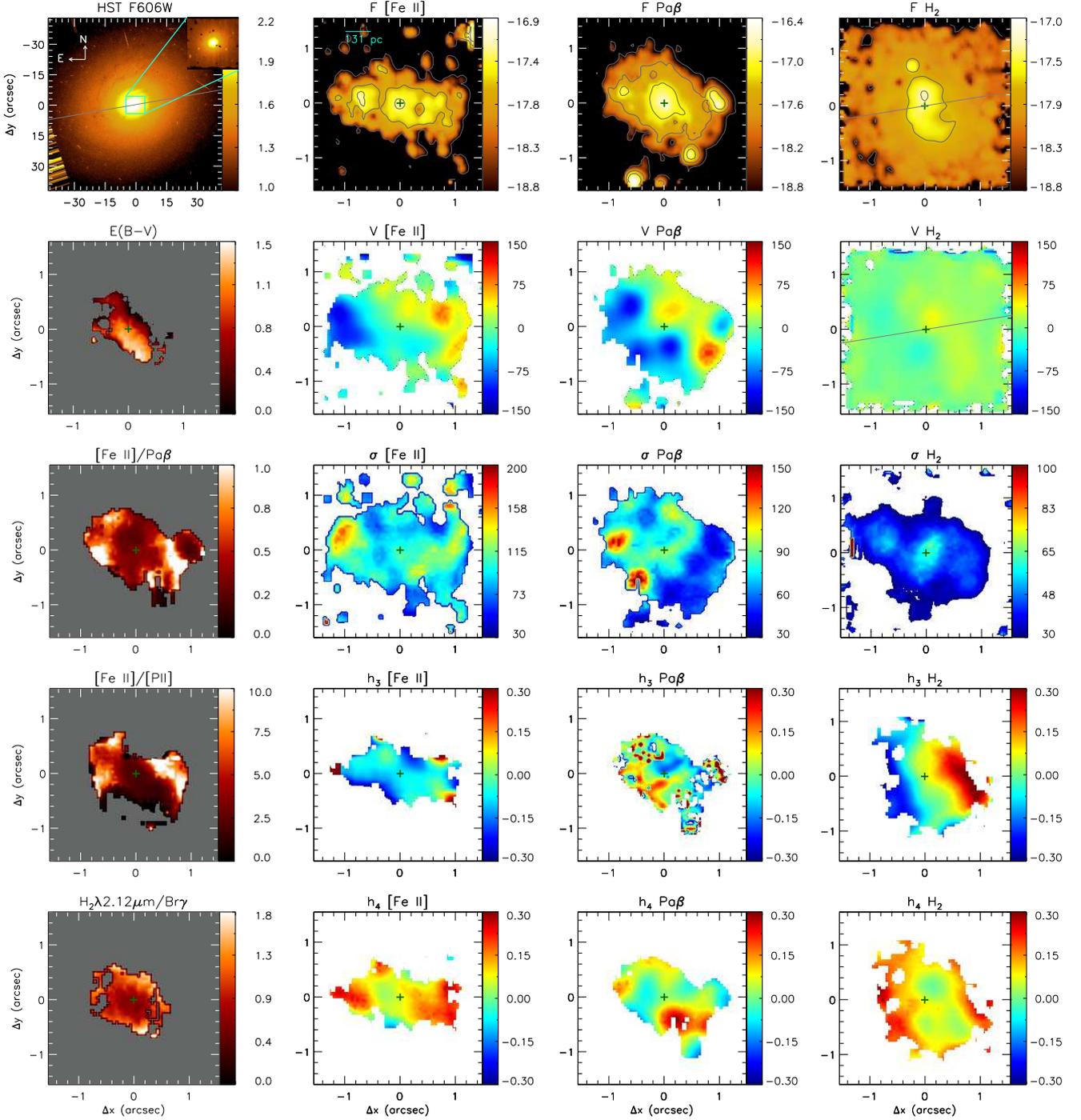}
\end{tabular}
\caption{Maps of properties derived from the emission-line profiles of NGC\,788 over the Gemini NIFS FOV (3$^{\prime\prime}\times$3$^{\prime\prime}$), shown as the small cyan square over the continuum image, in the top-left corner. First column, from top to bottom: HST-WFPC2 F606W continuum image \citep{malkan1998} and in the insert at the top right corner the NIFS J-band continuum image; the grey line marks the photometric major axis; 
reddening $E(B-V)$ map obtained from the Pa$\beta$/Br$\gamma$ line ratio; and line ratio maps identified on the top of each panel.
Second column, from top to bottom: Flux $F$, line-of-sight velocity V$_{\rm LOS}$, velocity dispersion $\sigma$, $h_3$ Gauss-Hermite and $h_4$ Gauss-Hermite moments for the [Fe\,{\sc ii}]$\lambda1.2570\mu$m emission line. Third and fourth columns: same as previous column for the Pa$\beta$ and H$_{2} \lambda2.1218\mu$m emission lines, respectively. Fluxes are shown in logarithmic units of erg s$^{-1}$ cm$^{-2}$, V$_{\rm LOS}$ in km\,s$^{-1}$, relative to the systemic velocity of the galaxy and $\sigma$ values are shown in km\,s$^{-1}$, after correction for the instrumental broadening.  This galaxy is individually discussed in Sec.\,\ref{sngc788}.}
\label{ngc788}
\end{figure*}

\begin{figure*}
\begin{tabular}{c}
\bf{\LARGE Mrk\,607} \\
\includegraphics[width=0.98\textwidth]{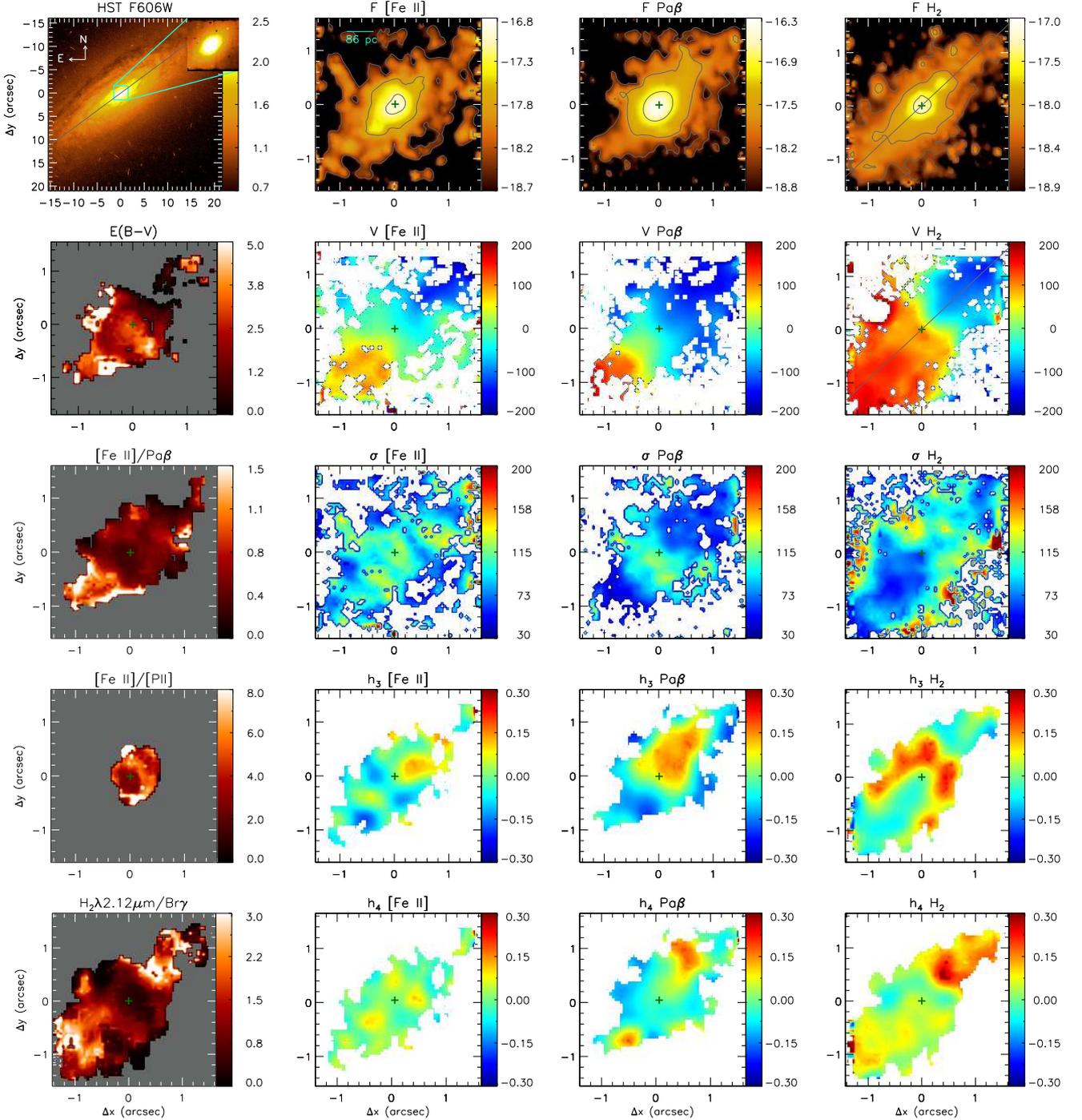}
\end{tabular}
\caption{As in Fig.\,\ref{ngc788} for Mrk\,607. This galaxy is individually discussed in Sec.\,\ref{smrk607}}
\label{mrk607}
\end{figure*}

\begin{figure*}
\begin{tabular}{c}
\bf{\LARGE NGC\,3227} \\
\includegraphics[width=0.98\textwidth]{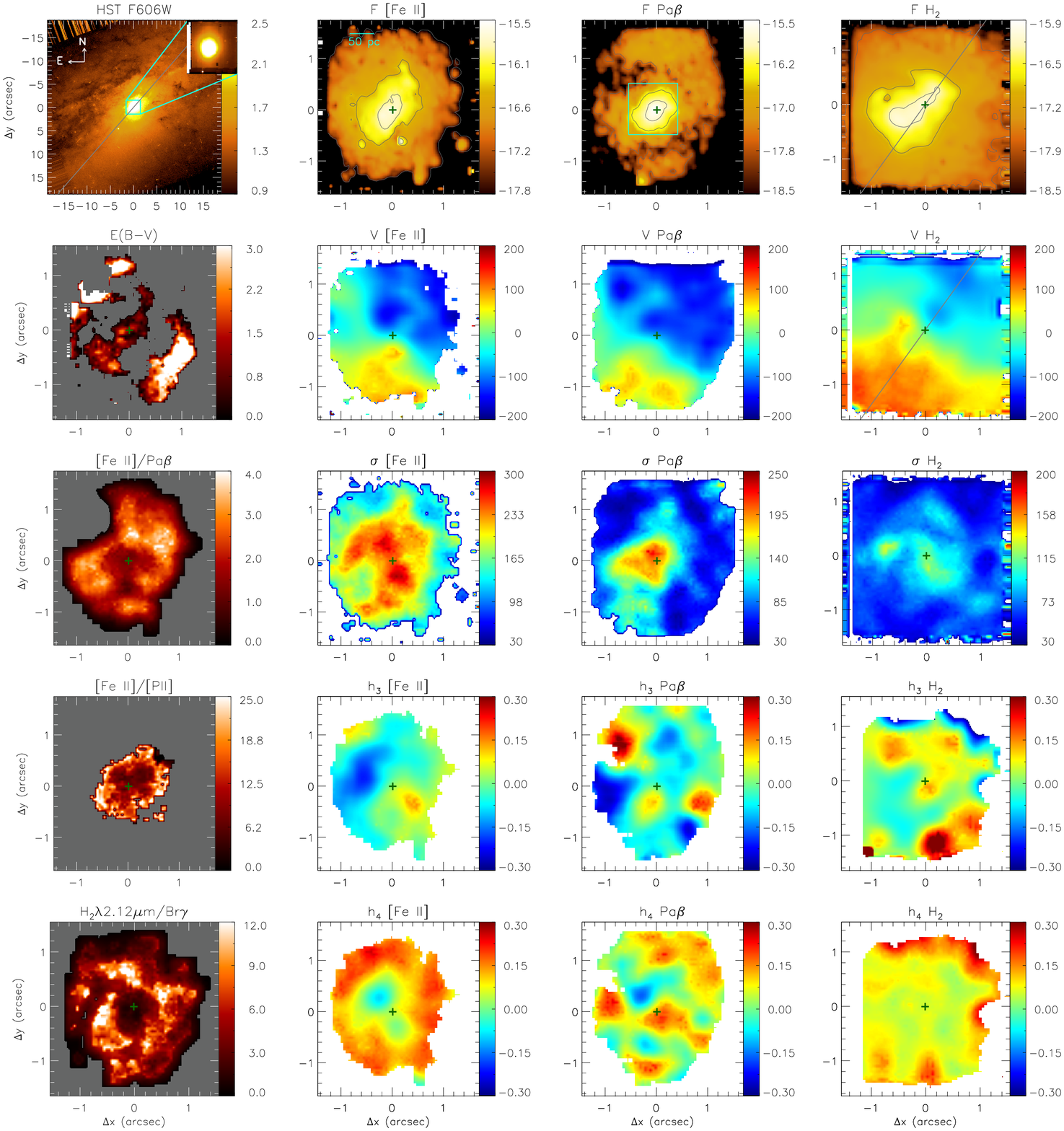}
\end{tabular}
\caption{As in Fig.\,\ref{ngc788} for NGC\,3227. The cyan square in the Pa$\beta$ flux map shows the region in which we had to subtract the contribution of a broad component. This galaxy is individually discussed in Sec.\,\ref{sngc3227}}
\label{ngc3227}
\end{figure*}

\begin{figure*}
\begin{tabular}{c}
\bf{\LARGE NGC\,3516} \\
\includegraphics[width=0.98\textwidth]{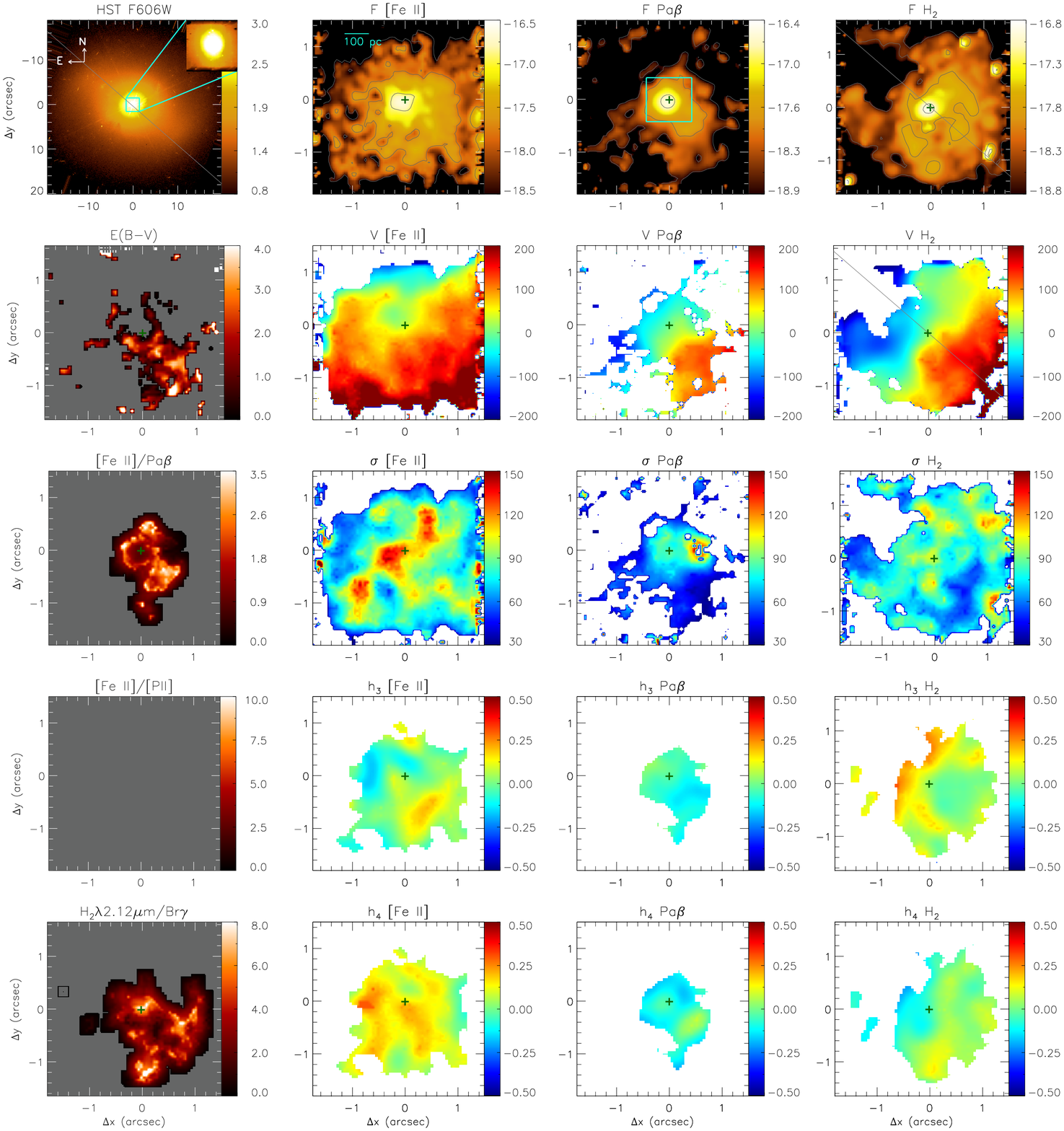}
\end{tabular}
\caption{As in Fig.\,\ref{ngc788} for NGC\,3516. The cyan square in the Pa$\beta$ flux map shows the region in which we had to subtract the contribution of a broad component. This galaxy is individually discussed in Sec.\,\ref{sngc3516}}
\label{ngc3516}
\end{figure*}

\begin{figure*}
\begin{tabular}{c}
\bf{\LARGE NGC\,5506} \\
\includegraphics[width=0.98\textwidth]{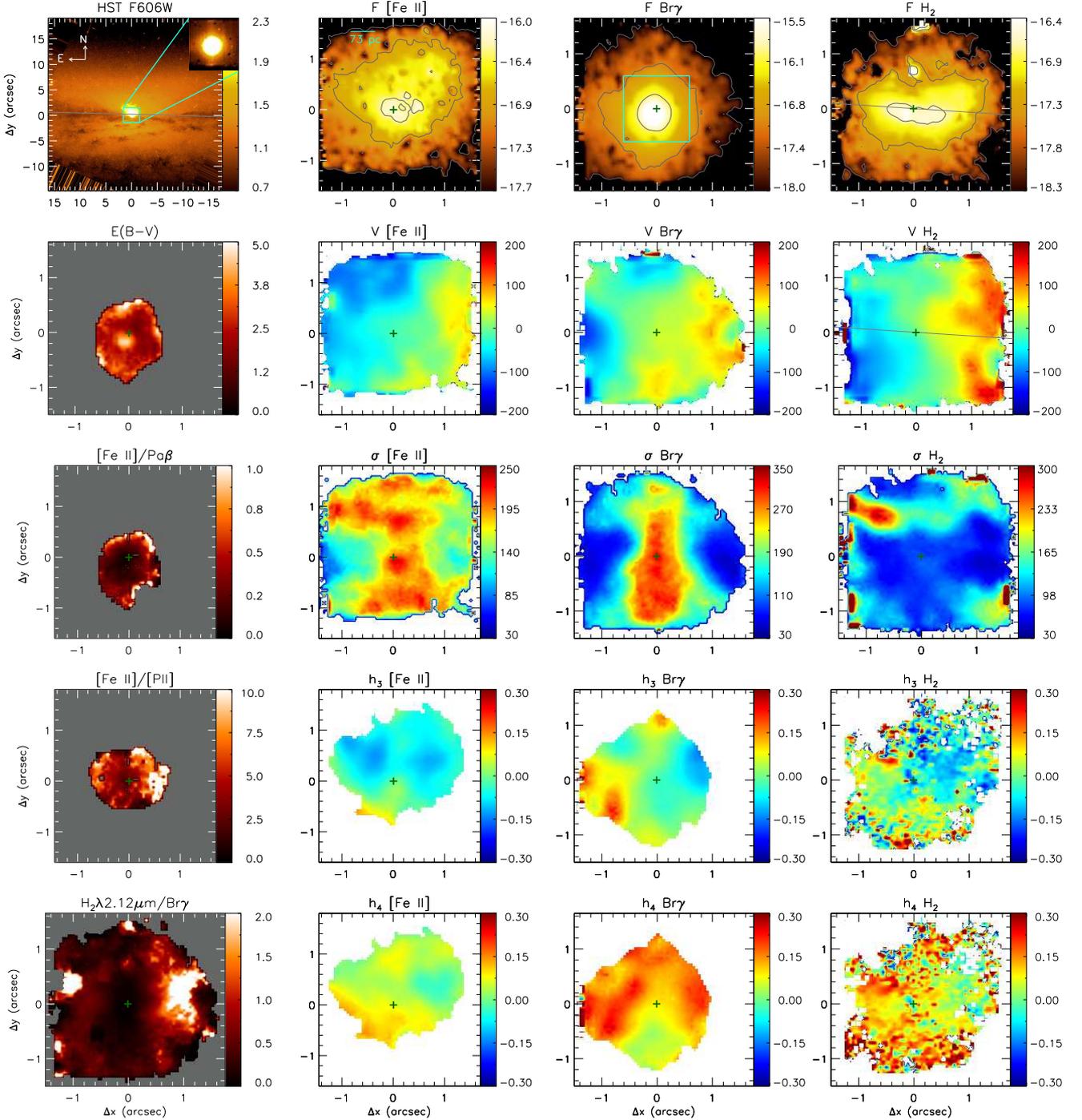}
\end{tabular}
\caption{As in Fig.\,\ref{ngc788} for NGC\,5506. The cyan square in the Pa$\beta$ flux map shows the region in which we had to subtract the contribution of a broad component (although less broad than in the previous Seyfert 1 galaxies). This galaxy is individually discussed in Sec.\,\ref{sngc5506}}
\label{ngc5506}
\end{figure*}

\begin{figure*}
\begin{tabular}{c}
\bf{\LARGE NGC\,5899} \\
\includegraphics[width=0.98\textwidth]{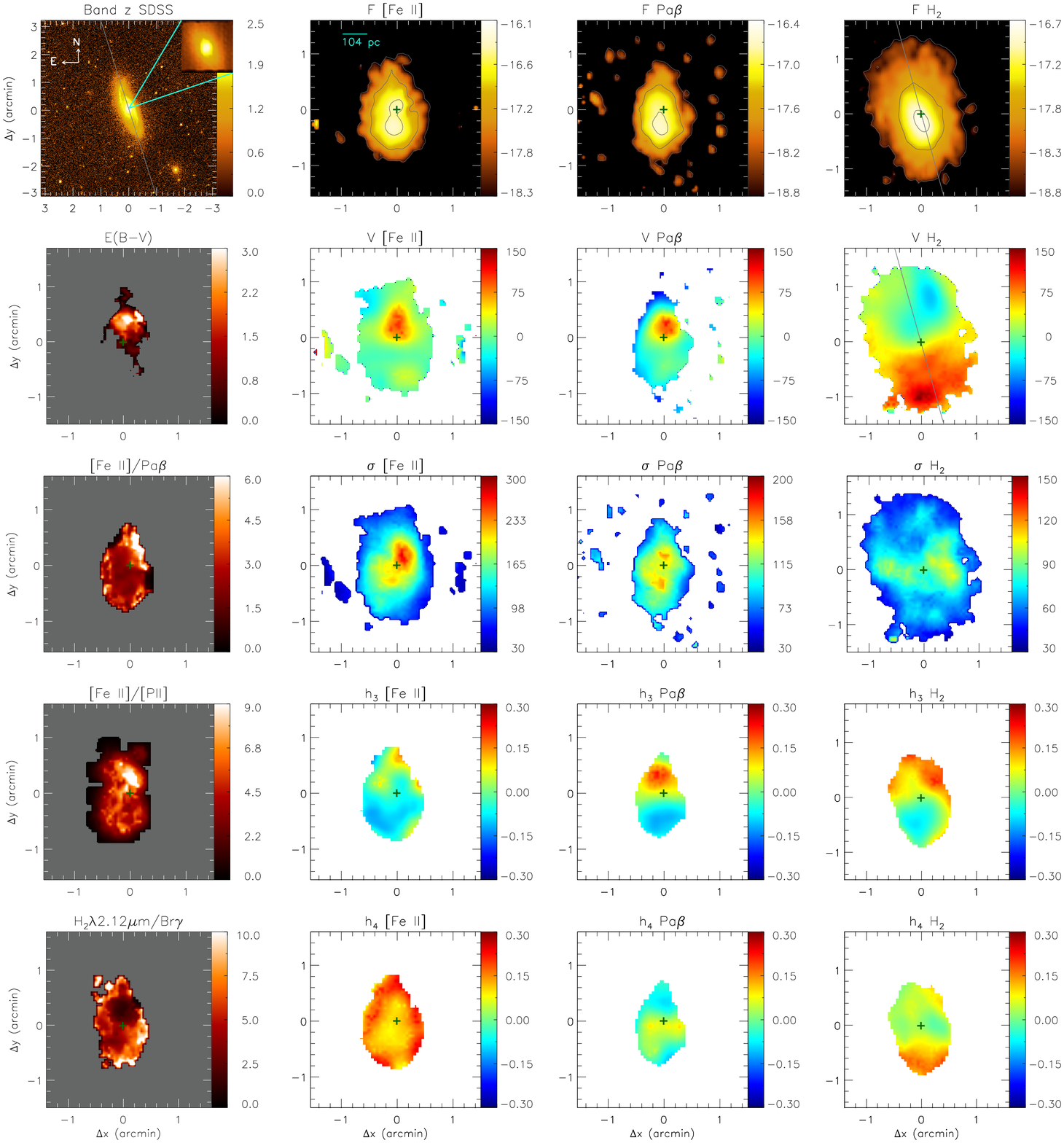}
\end{tabular}
\caption{As in Fig.\,\ref{ngc788} for NGC\,5899, but with the large-scale image from SDSS at the z-band \citep{baillard2011}. This galaxy is individually discussed in Sec.\,\ref{sngc5899}.}
\label{ngc5899}
\end{figure*}

From the fits of the line profiles and resulting parameters we have constructed maps of: flux distributions, V$_{\rm LOS}$, velocity dispersions, $h_3$ and $h_4$ moments as well as of the reddening $E(B-V)$ and emission line ratio maps, presented in Fig.\,\ref{ngc788} for NGC\,788, Fig.\,\ref{mrk607} for Mrk\,607, Fig.\,\ref{ngc3227} for NGC\,3227, Fig.\,\ref{ngc3516} for NGC\,3516, Fig.\,\ref{ngc5506} for NGC\,5506 and Fig.\,\ref{ngc5899} for NGC\,5899. We chose to show the results only for [Fe\,{\sc ii}]$\lambda$1.2570$\mu$m, Pa$\beta\lambda$1.2822$\mu$m (or Br$\gamma \lambda$2.1661$\mu$m in one case) and H$_{2} \lambda$2.1218$\mu$m because these are the strongest emission lines allowing the mapping of the ionized gas properties via the [Fe\,{\sc ii}] and Pa$\beta$ (or Br$\gamma$) emission lines, and the warm molecular gas properties through the H$_2$ emission line above. Although the nuclear spectra of some galaxies show further coronal lines, as [S\,{\sc ix}]$\lambda$1.2525\,$\mu$m and  [Ca\,{\sc viii}]$\lambda$2.3211\,$\mu$m, their flux distributions are unresolved or barely resolved and are not shown here, leaving a detailed discussion about these lines together with that of the gas excitation to be presented in Paper B.

In the next subsections we discuss separately the results for each map, where we have masked out pixels with bad line fits, flagged according to the {following criteria, applied to all maps: (1)  relative uncertainties in the line fluxes larger than 30\%; (2) uncertainties in velocity and velocity dispersions larger than 50 km\,s$^{-1}$.}


\subsection{Flux Distributions}

Flux maps are shown in the first row of Figs.\,\ref{ngc788} -- \ref{ngc5899}, where we have drawn the Position Angle (PA) of the major axis of the galaxy as given in the HYPERLEDA database \citep{hyper} over the continuum images (leftmost panel) and H$_2$ flux and velocity maps. In all emission-line maps, the peak flux is observed at the location of the continuum peak, defined as the origin of the coordinates and identified with the galaxy nucleus. The flux distributions for the selected emission lines cover most of the FOV, corresponding to maximum distances from the nucleus varying from 100\,pc for NGC\,3227 to 410\,pc for NGC\,788.

The direction of the largest extent of the [Fe\,{\sc ii}]$\lambda1.25\,\mu$m and Pa$\beta$ flux distributions follows the orientation of the ionization axis, and which in many cases approximately coincides with the orientation of the major axis of the galaxy in our sample. Exceptions are the cases of NGC\,788 where they are oriented at a small angle relative to the major axis, and NGC\,5506, where the largest extent of the [Fe\,{\sc ii}] emission is perpendicular to the major axis. 
Although also showing the highest emission levels following the major axis orientation, the H$_2\lambda2.12\,\mu$m flux maps usually show emission more spread over all directions, being less collimated than the ionized gas emission. 
The only exception is NGC\,788, for which the strongest H$_2$ emission is more extended perpendicularly to the major axis of the galaxy.

\subsection{Line-ratio maps}

Here we describe briefly the main features observed in the line-ratio maps; a more in-depth discussion of the gas excitation will be presented in Paper B.

The reddening and line-ratio maps are shown in the first column of Figs.\,\ref{ngc788} -- \ref{ngc5899}, below the galaxy continuum image.

The $E(B-V)$ map was obtained from:
\begin{equation}
E(B-V)=4.74\log\left(\frac{5.88}{F_{\rm Pa\beta}/F_{\rm Br\gamma}}\right),
\end{equation}
where $F_{\rm Pa\beta}$ and $F_{\rm Br\gamma}$ are the corresponding line fluxes {and} we have adopted in the derivation of the expression above the \citet{cardelli} extinction law and $F_{\rm Pa\beta}/F_{\rm Br\gamma}$ theoretical ratio of $5.88$ \citep{osterbrock}.

The $E(B-V)$ values are mostly in the range 1--3, reaching the highest values at the nucleus of NGC\,5506 ($E(B-V)\approx 5$) and NGC\,5899 ($E(B-V)\approx3$) while for the other galaxies the highest values are observed outwards.

The [Fe\,{\sc ii}]$\lambda$1.2570$\mu$m/Pa$\beta$ ratio maps can be used to investigate the excitation mechanism of [Fe\,{\sc ii}] \citep[e.g.]{ardila05,riffel08b,rogerio09,storchi2009}, with typical values for Seyfert galaxies ranging from 0.6 to 2. Most values are observed within this range, {usually} increasing from the centre outwards. In the case of NGC\,3227 its value increases to $\approx$\,4 in a region with the shape of an off-centered ring surrounding the nucleus in which an enhanced velocity dispersion in the [Fe\,{\sc ii}] emission lines is observed. Higher values than 2 are also observed in NGC\,3516 and NGC\,5899, also in association with regions of {higher} velocity dispersion than the surroundings, suggesting the contribution of shocks for the gas emission.

The [Fe\,{\sc ii}]$\lambda$1.2570$\mu$m/[P\,{\sc ii}]$\lambda$1.8861$\mu$m line ratio can also be an indicator of shocks \citep{storchi2009}, if {its} value becomes larger than $\approx$\,2. The corresponding maps are compact, due to the small extent of the [P\,{\sc ii}] emission, with typical values of $\approx$5 at the nucleus and increasing to $\approx$10 at $\approx 0\farcs5$, thus supporting the presence of shocks. In the cases of NGC\,3227 and NGC\,5899, the off-nuclear values reach values larger than 10, also in association with high [Fe\,{\sc ii}] $\sigma$ values, supporting even stronger contribution from shocks.

The H$_2\,\lambda$2.1218$\mu$m/Br$\gamma$, used to investigate the origin of the H$_2$ excitation, is usually in the range 0.6 to 2 for Seyfert galaxies \citep{ardila05,riffel08b,rogerio09,storchi2009}. Most measured values are indeed in this range, with a general behavior of showing the lowest {ratios} at the nucleus and  increasing outwards, that we attribute to the destruction of the H$_2$ molecule by the strong radiation field close to the nucleus. The lowest values -- lower than 2 -- are observed for NGC\,788 and NGC\,5506, while the highest, reaching values in the range 8--10, are seen again in NGC\,3227, NGC\,3516 and NGC\,5899, at similar locations as those where an enhancement is seen in [Fe\,{\sc ii}]/Pa$\beta$ and [Fe\,{\sc ii}] $\sigma$ values, supporting again a contribution from shocks.

\subsection{Line-of-sight Velocity Fields} 

The line-of-sight velocity (V$_{\rm LOS}$) fields are shown in the second line of Figs.\,\ref{ngc788} -- \ref{ngc5899}, after the subtraction of the  systemic velocity obtained through the fit of a rotating disk model to the H$_2$ velocity field. Again, here we just describe the general features of these velocity fields, as a proper discussion is deferred to Paper B.

The most common characteristic of the velocity fields is the rotation pattern, that is clearer in the H$_2$ velocity maps (presenting the typical ``spider diagram'' structure \citep{gal_dynamics-book}). The H$_2$ velocity amplitudes is similar to those also seen in the [Fe\,{\sc ii}] and Pa$\beta$ velocity fields, except in the case of NGC\,788, for which the H$_2$ rotation amplitude is smaller. The rotation pattern can also be seen in the [Fe\,{\sc ii}] and Pa$\beta$ velocity fields, but in these cases it is usually disturbed -- more in the case of [Fe\,{\sc ii}] than in the case of Pa$\beta$ -- indicating the presence of additional kinematic components, usually associated with enhanced $\sigma$ values.

\subsection{Velocity dispersion maps}

These maps are shown in the third line of Figs.\,\ref{ngc788} -- \ref{ngc5899}.

The [Fe\,{\sc ii}] $\sigma$ maps usually show the highest values, reaching up to $\approx$150--300 km\,s$^{-1}$, while the H$_2$ maps show the lowest values, in the range $\approx$40--90\,km\,s$^{-1}$. The Pa$\beta$ $\sigma$ maps are similar to those of [Fe\,{\sc ii}] although reaching somewhat overall lower values, in the range $\approx$40\,km\,s$^{-1}$ to 250\,km\,s$^{-1}$. An exception is the case of NGC\,5506 that shows regions of enhanced $\sigma$ to $\approx$ 350 km\,s$^{-1}$) for all emission lines, extended perpendicularly to the major axis for [Fe\,{\sc ii}] and Pa$\beta$ and in patches parallel to the major axis in H$_2$; such patches are also seen in [Fe\,{\sc ii}]. 

\subsection{ $h_3$ and  $h_4$ Gauss-Hermite moments}
The $h_3$ and $h_4$ maps show values ranging from --0.3 to 0.3 for all emission lines. An inverse correlation between the $h_3$ map and the velocity fields is observed for all emission-lines: positive values of $h_3$ are seen at the locations where negative velocities are observed in the velocity fields, while negative values are seen where positive velocities are observed. This means that red wings are observed in blueshifted emission lines, while blue wings are observed in redshifted emission lines. 


We also observe in some cases (e.g. for the three emission lines in NGC\,3227 and some of the lines in the other galaxies) an inverse correlation between the $h_4$ and the velocity dispersion maps. Positive values and highest values of $h_4$ are seen at the locations showing the lowest velocity dispersion values, while values zero or most negative ones are seen in the regions with the highest velocity dispersion.

\subsection{Total gas mass and surface mass density distributions}

The measurement of gas masses and surface mass densities within the inner few hundred pc of the host galaxies of AGN can be used to evaluate the gas reservoir available to trigger and maintain the nuclear activity, as well as the formation of new stars in the circumnuclear region. The presence of recent star formation in the circumnuclear region of active galaxies has been evidenced via the observation of low stellar velocity dispersion structures (Riffel et al. 2017) and their association with young stars \citep[e.g.][]{storchi2012}. We can also use gas mass estimates in the vicinity of AGN to look for a possible correlation between the amount of available gas and the power of the AGN.

We have calculated the ionized gas masses, in units of solar masses (M$_{\odot}$) using the following expression \citep[e.g.][]{scoville,riffel08,storchi2009,astor2014,astor2017}:

\begin{equation}
\label{mhii}
M_{\rm HII} \approx 5.1 \times 10^{16}\left(\frac{F_{\rm Pa\beta}}{\rm erg\,s^{-1}cm^{-2}}\right) \left(\frac{D}{\rm Mpc}\right)^2 [M_\odot] ,
\end{equation}

\noindent where $F_{\rm Pa\beta}$ is the flux in the Pa$\beta$ line and $D$ is the distance to the galaxy.    

This equation was obtained from \citet{scoville} assuming an electron temperature of $T=10^4$\,K, electron density of $N_e=100\,$cm$^{-3}$ and case B recombination \citep{osterbrock}, values applicable to the inner kiloparsec of active galaxies. The flux of Br$\gamma$ used in the original equation was replaced by the stronger Pa$\beta$ one assuming the theoretical ratio $F_{\rm Pa\beta}$/$F_{\rm Br\gamma}$=5.88 \citep{osterbrock}.{For NGC\,5506, we did use the flux of Br$\gamma$ emission line instead of Pa$\beta$, as the signal to noise in the J-band spectrum is worse than in the K-band one, particularly in the extranuclear regions.}

The mass of the warm molecular gas (in M$_{\odot}$) was obtained using \citep{scoville}: 

\begin{equation}
\label{mh2}
M_{\rm H_2} \approx 5.0776 \times 10^{13} \left(\frac{F_{\rm H_{2}\lambda2.1218}}{\rm erg\,s^{-1}cm^{-2}}\right) \left(\frac{D}{\rm Mpc}\right)^2 [M_\odot], 
\end{equation}
where $F_{\rm H_{2}\lambda2.1218}$ is the flux for the corresponding emission line. 

In the derivation of the above equation by \citet{scoville} it was assumed that the vibrational temperature is $T=2000$\,K, indicating a thermalized gas which is valid for $n_{\rm H_2}>10^{4.5}$\,cm$^{-3}$. Previous studies by members of our group have found that the assumption of thermal equilibrium is indeed applicable for nearby AGNs and the derived temperatures are very close to the above value \citep{storchi2009,diniz2015}.

 Uncertainties in the derived mass values were obtained considering that the mass derived from both  Eq.~\ref{mhii} and \ref{mh2} are directly proportional to the line fluxes. All the mass uncertainties are thus lower than 30\%, as we have masked out regions with flux uncertainties larger than this value. The actual uncertainties may be somewhat higher due to further uncertainties in the absolute flux calibration. In addition, we cannot derive precise physical parameters such as gas density and temperature, thus, we have to rely on the above assumptions even though they are just ``educated estimates" based on our previous similar studies.



\begin{figure*}
\includegraphics[width=0.9\textwidth]{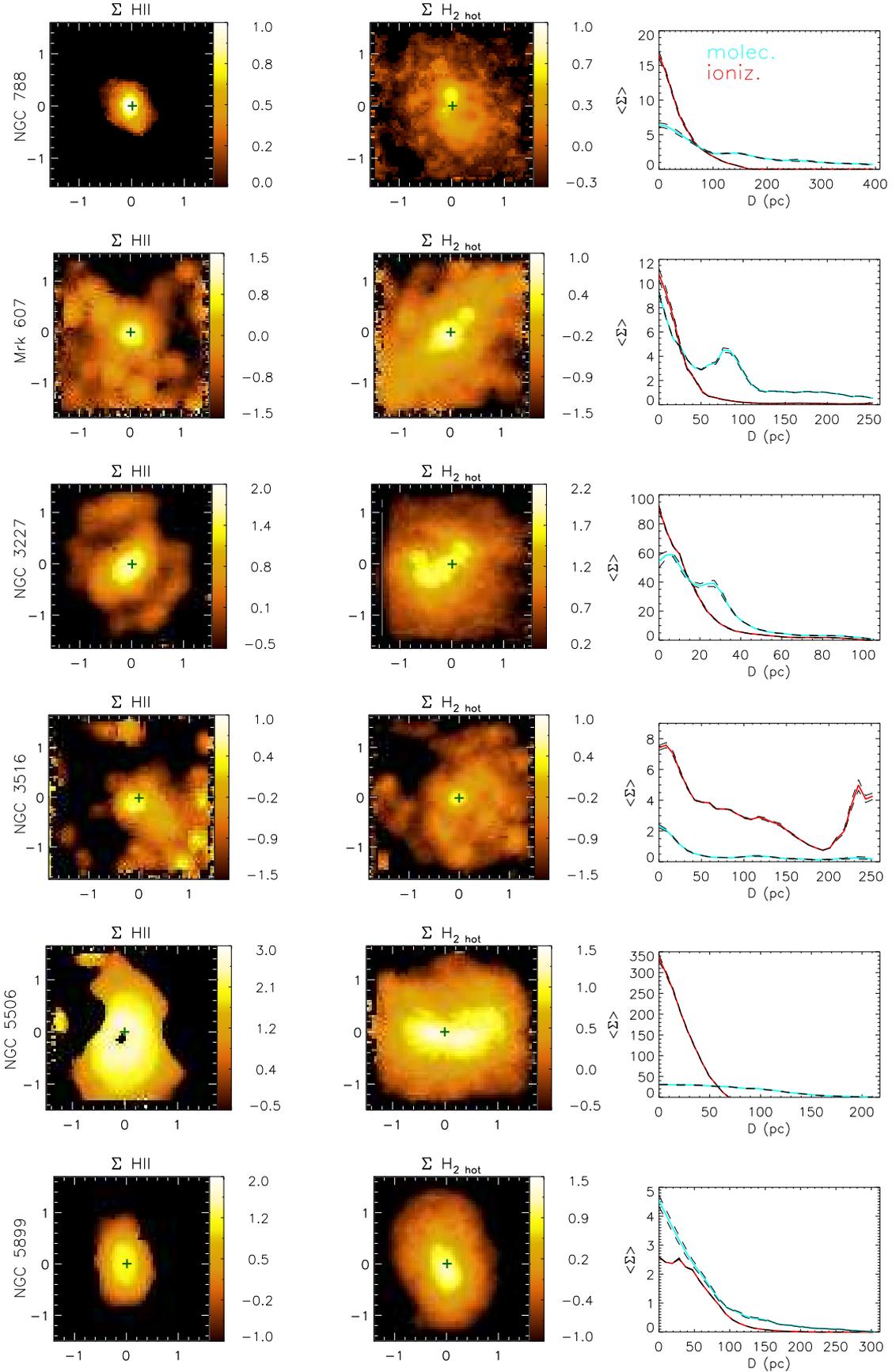}
\caption{Surface gas mass density distributions $\Sigma$: in the first column we show $\Sigma_{\rm HII}$ in units of M$_{\odot}$ pc$^{-2}$); in the second column $\Sigma_{\rm H_2}$ in units of 10$^{-3}$ M$_{\odot}$ pc$^{-2}$, and in the third the corresponding azymuthaly averaged gradients, showing more centraly peaked distributions for the ionized gas.}
\label{mass}
\end{figure*}

We show the surface mass density $\Sigma$ distributions of ionized and warm molecular gas -- $\Sigma_{\rm HII}$ and $\Sigma_{\rm H_2}$, respectively -- in the first two columns of panels of Fig.\,\ref{mass}, and the corresponding azymuthally averaged profiles in the third column, in units of solar masses per parsec square. The surface mass densities were obtained by calculating the gas masses in each spaxel using the equations above and dividing them by the surface area of each spaxel, taking into account projection effects via the use of the disk inclinations quoted in \citet{riffel2018}. The resulting ionized gas surface mass density  distributions are in some cases more compact when compared with that of the warm molecular gas, that seems to extend farther and more uniformly from the nucleus than the ionized gas.

The azimuthaly-averaged profiles of $\Sigma_{\rm HII}$ and $\Sigma_{\rm H_2}$, shown in the last column of Fig.\,\ref{mass}, reveal that typical ratios between the two surface mass densities within the inner $\approx$\,100\,pc  are of the order of 10$^3$ and that $\Sigma$ gradients are steeper for the ionized than for the molecular gas.


\begin{table*}
\begin{small}
\caption{{Areas, masses and average surface mass density $\Sigma$ values for the ionized, warm and estimated cold molecular gas within the Gemini NIFS FOV.}} 
\vspace{0.3cm}
\begin{tabular}{l c c c c c c c c c c c c l}
\hline
 Galaxies   & Area (H$_2$) & Area (H{\sc ii}) & M (H$_{2}$)$_{\rm hot}$  & M (H$_{2}$)$_{\rm cold}$  & M (H{\sc ii}) & $\Sigma$ (H$_{2}$)$_{\rm hot}$ & $\Sigma$ (H$_{2}$)$_{\rm cold}$  & $\Sigma$ (H{\sc ii}) \\
 & 10$^5$ pc$^2$ & 10$^5$ pc$^2$ & 10$^2$ M$_{\odot}$  & 10$^7$ M$_{\odot}$  & 10$^5$ M$_{\odot}$  & 10$^{-3}$ M$_{\odot}$/pc$^2$  &  10$^{3}$ M$_{\odot}$/pc$^2$  & M$_{\odot}$/pc$^2$ \\
\hline
NGC\,788  & 6.2 & 5.5 & 5.9$\pm$0.15 & 43$\pm$1 & 28$\pm$9.8 & 0.9$\pm$0.03 & 0.8$\pm$0.02 & 5.2$\pm$1.8  \\
Mrk\,607  & 0.6 & 0.2 & 1.1$\pm$0.2 & 8$\pm$1 & 6.4$\pm$4.9 & 1.8$\pm$0.3 & 3.3$\pm$0.6 & 28.0$\pm$22.0  \\
NGC\,3227 & 0.9 & 0.4 & 7.5$\pm$0.4 & 54$\pm$3 & 10.0$\pm$1.0 & 8.0$\pm$0.4 & 12.0$\pm$0.6 & 23.0$\pm$2.0  \\
NGC\,3516 & 0.8  & 0.6   & 3.0$\pm$1.7 & 22$\pm$12 & 1.8$\pm$1.0  & 3.7$\pm$2.0 & 3.6$\pm$2.0 & 2.9$\pm$1.6 \\
NGC\,5506 & 1.4 & 1.2 & 8.2$\pm$0.4 & 60$\pm$3 & 190.0$\pm$10.0 & 5.7$\pm$0.3 & 4.6$\pm$0.2 & 140.0$\pm$7.0  \\
NGC\,5899 & 1.9 & 0.9 & 3.3$\pm$0.3 & 23$\pm$2 & 2.5$\pm$1.8 & 1.7$\pm$0.2 & 2.7$\pm$0.2 & 2.8$\pm$2.0  \\

 \hline
\end{tabular}
 \label{tabmass}
\end{small}

\end{table*}

In Table\,\ref{tabmass} we show the integrated gas mass values (for the whole FOV) as well as the average surface mass densities (in units of M$_{\odot}$ pc$^{-2}$) obtained as the ratio between the integrated masses and the area over which 
they are distributed, listed also in the Table.   
We also include in Table\,\ref{tabmass} an estimate of the cold molecular gas mass. A number of studies have derived the ratio between the cold and warm H$_2$ gas masses by comparing the masses obtained using the cold CO molecular lines observed in millimetric  wavelenghts with that of the warm H$_2$ observed in the near-IR. \citet{dale05} obtained ratios in the range  $10^5$--10$^7$, while \citet{ms2006}, using a larger sample of 16 luminous and ultra-luminous infrared galaxies,  derived a ratio M$_{\rm cold}$/M$_{\rm warm}$ = 1\,--\,5\,$\times\,10^6$. More recently, \citet{ma2013} compiled from the literature values of M$_{\rm cold}$ derived from CO observations and H$_2$(2.12$\mu$m) luminosities for a larger number of galaxies, covering a wider range of luminosities, morphological and nuclear activity types. From those data, they propose that an estimate of the cold H$_2$ gas mass can be obtained from the flux of the warm H$_2\lambda$2.1218 line as:

\begin{equation}
M_{\rm H_2\,{cold}} \approx 1174\,\left(\frac{L_{\rm H_2\lambda2.1218}}{L_{\odot}}\right),
\end{equation}
{where $L_{\rm H_2\lambda2.1218}$ is the luminosity of the H$_2$(2.12$\mu$m) line. The error associated to the factor $\beta=$ 1174 is $\approx$ 35\%, according to \citet{ma2013}.}


Table \ref{tabmass} shows that the integrated mass of warm H$_2$ gas (within typically 100 -- 400 pc from the nucleus) ranges from 106 M$_{\odot}$ for Mrk\,607 to 820\,M$_{\odot}$ for NGC\,5506, while the estimated mass of cold molecular gas ranges from 8$\times$10$^7$M$_{\odot}$ for Mrk\,607 to 6$\times$10$^{8}$ M$_{\odot}$ for NGC\,5506. The masses of ionized gas range from 1.8$\times$10$^5$ M$_{\odot}$ for NGC\,3516 to 1.9$\times$10$^7$  M$_{\odot}$ for NGC\,5506. We point out again that the uncertainties in the calculated masses, presented in Tab.\,\ref{tabmass}, were obtained only by propagating the line flux uncertainties and are thus lower limits, considering that there are additional uncertainties  in adopted physical properties -- e.g. on the $M_{\rm H_2\,{cold}}/M_{\rm H_2}$ ratio, H$_2$ vibrational temperatures and electron densities.

\section{Discussion}
\label{sec:discussion}
In this section we discuss individually each galaxy {presenting a brief review of previous results from the literature, trying to relate them with ours}. We also {discuss and analyze the masses of ionized and molecular gas, as well as their surface mass density distributions and the global properties of the gas kinematics, leaving the kinematic modelling and the analysis of the gas kinematics and excitation to Paper B}. 

\subsection{NGC\,788}
\label{sngc788}

NGC\,788 (Fig.\,\ref{ngc788}) was identified as a Seyfert galaxy by \citet{huchra1982}, and has been observed in the optical \citep{hamuy1987,wagner1988,kay1994,cruz1994}, radio \citep{ulvestad1989}, and
millimetre \citep{heckman1989} wavebands. This galaxy is an early-type spiral with faint arms visible up to $\approx$ 30\farcs from the nucleus \citep{evans1996}, most conspicuous to the north-west of the nucleus with a string of {bright compact} H{\sc  ii} regions, while a complex of fainter H{\sc  ii} regions is associated with a southern arm \citep{evans1996}.

The kinematics major axis PA\,$\approx$\,125$^{\circ}$  was determined from the $V_{\rm LOS}$ fields (Fig.\,\ref{ngc788}), with a value close to that obtained by \citet{riffel2017} for the stellar kinematics of $\approx$ 130$^{\circ}$.
The velocity amplitude observed for [Fe\,{\sc ii}] and Pa$\beta$, of $\approx$\,150\,km\,s$^{-1}$ is much higher than that for H$_2$, of only $\approx$\,50\,km\,s$^{-1}$. These higher velocities are associated {with a more collimated emission and higher velocity dispersions suggesting the presence of a bipolar outflow along the east-west direction, as {suggested from the morphology of the gas emission} in the [Fe\,{\sc ii}] flux map.}


The $E(B-V)$ map from Fig.\,\ref{ngc788} has the highest values to southwest of the nucleus, which coincide with a dust lane seen in the HST - F606W optical continuum image from \cite{martini2003}.  

\subsection{Mrk\,607}
\label{smrk607}
Mrk\,607 (Fig.\,\ref{mrk607}) is an Sa galaxy hosting a Seyfert 2 nucleus. A continuum image from \citet{ferruit2000} {shows a high ellipticity of $\approx$ 0.60, yielding an inclination i$\approx$ 67$^{\circ}$), with major axis PA of $-43^{\circ}$, which agrees with our estimate of the kinematic major axis orientation} of $\approx$ $-38^{\circ}$ from the H$_2$ velocity field (Paper B).

Our data reveal that the three velocity fields -- [Fe\,{\sc ii}], Pa$\beta$ and H$_2$ -- show what seems to be a rotation pattern, very steep in the center in H$_2$ and less steep in the first two lines. Spiral dust lanes are clearly visible in the HST F547 continuum image of the inner 8$^{\prime\prime}$ (1.4 kpc) \citep{ferruit2000}. The high inclination of the galaxy may explain the large velocity dispersions observed. In the central 5$^{\prime\prime}$ (875 pc) it presents weak radio emission extended along PA $\approx$ 180$^{\circ}$ \citep{colbert1996,nagar1999} which is not aligned with the axis of the most extended gas emission, which is observed along the galaxy major axis. 

The $E(B-V)$ map from Fig.\,\ref{mrk607} has high values ($\approx$ 5) to southeast/west of the nucleus, which coincide with a dust lane seen in the HST - F606W optical continuum image from \cite{martini2003}. 

\subsection{NGC\,3227}
\label{sngc3227}
NGC\,3227 (Fig.\,\ref{ngc3227}) is a well studied barred galaxy, with a Seyfert 1.5 nucleus \citep{ho1997} and in interaction with the dwarf elliptical galaxy NGC\,3226 \citep{mundell2004}. The galactic disc has an inclination of 56$^{\circ}$, with an outer photometric major-axis at a position angle PA = 158$^{\circ}$ \citep{mundell1995}. This value agrees with the PA$\approx$ 155$^{\circ}$ we have obtained from the fit of the H$_2$ V$_{\rm LOS}$ velocity field (Paper B), and with the value of 156$^{\circ}$ found by \citet{riffel2017} from the stellar kinematics. The three emission-line velocity fields show a rotation pattern, that is nevertheless distorted due to the presence of additional components, mainly to the north-east, east and south-east. These components are most conspicuous in [Fe\,{\sc ii}] and Pa$\beta$ velocity maps and are associated with regions of enhanced velocity dispersion.     
 
The central region has been mapped in $^{12}$CO(1-0) and $^{12}$CO(2-1) by \citet{schinnerer2000}, who detected molecular gas very close to the nucleus (within $\approx$ 13 pc), in agreement with our results that show a large molecular gas concentration towards the centre (see the surface mass density profile in the third line of Fig.\,\ref{mass}). In addition, \citet{schinnerer2000} found an asymmetric nuclear ring with a diameter of about 3\farcs0 (220 pc), which seems to correspond to the structure delineated in the line ratio maps [Fe\,{\sc ii}]/Pa$\beta$ and H$_2$/Br$\gamma$ of Fig.\,\ref{ngc3227}.

The inner kiloparsec hosts a radio jet at PA $\approx - 10^{\circ}$ and a conical NLR outflow at PA\,$\approx$\,15$^{\circ}$ \citep{mundell1995}, while \citet{arribas1994} report an H$\alpha$ outflow at PA\,$\approx$\, 50$^{\circ}$. The extent of the [Fe\,{\sc ii}] emission to the north, as well as its enhanced velocity dispersion may be related to the radio emission, while the outflows observed at PAs between {15$^{\circ}$} and 50$^{\circ}$ can explain the deviation from rotation observed in the velocity fields of [Fe\,{\sc ii}] and Pa$\beta$ to the north-east. 
Signatures of gas outflows in NGC\,3227 have also been observed by \citet{davies2014} in the inner $1^{\prime\prime}-2^{\prime\prime}$ using near-IR IFS observations with the instrument SINFONI at the VLT. A $V-H$ colour map shows dust lanes to the south-west of the nucleus \citep{martini2003,davies2014}, which are co-spatial with the highest  $E(B-V)$ values seen in Fig.\,\ref{ngc3227}.


\subsection{NGC\,3516}
\label{sngc3516}
The Seyfert 1 nucleus of this SB0 galaxy (Fig.\,\ref{ngc3516}) shows variable ultraviolet absorption lines \citep{voit1987,kriss1996} as well as variable broad emission lines and continuum \citep{koratkar1996}. Spectroscopic studies of the gaseous kinematics using either long-slit \citep{ulrich1980,goad1987,goad1988,mulchaey1992} or integral field spectroscopy \citep{veilleux1993,aoki1994,arribas1994} have outlined multiple spectral components displaying strong deviations from ``normal'' galaxy rotation. Two general models have been proposed to explain the morphology and the kinematics of the emission-line gas that presents a Z-shaped structure covering the inner 5\farcs0: a bent bipolar mass outflow, first suggested by \citet{goad1987}
and further developed by \citet{mulchaey1992} and \citet{veilleux1993}, and a precessing twin-jet model by \citet{veilleux1993}.
The stellar rotation curve obtained by \citet{arribas1997} yields a systemic velocity of 2593$\pm$15 km\,s$^{-1}$, close to that derived by \citet{riffel2017} of $\approx$ 2631 km\,s$^{-1}$. The kinematic major axis obtained by \citet{arribas1997} has PA= 53$^{\circ}\pm5^{\circ}$ and is consistent with our results for the stellar kinematics \citep{riffel2017} and also with the orientation of the kinematic major axis of the molecular gas of $\approx$ 60$^{\circ}$.

\subsection{NGC\,5506} 
\label{sngc5506}
The nucleus of NGC\,5506 (Fig.\,\ref{ngc5506}) is classified as a Sy1.9 based on the detection of broad wings in the Pa$\beta$ profile \citep{blanco1990}. However, more recently, \citet{nagar2002} presented evidence that NGC\,5506 is an obscured narrow-line Sy1, via the detection of the permitted O\,{\sc i}$\lambda$1.129$\mu$m line, together with a broad pedestal of Pa$\beta$ and rapid X-ray 
variability. The galaxy nucleus is very compact in the mid-IR, with an apparent optical depth of $\tau^{app}_{10\mu m}$ $\approx$ 1.4 \citep{roche1991,roche2007,siebenmorgen2004}, although NGC\,5506 also shows variations in its silicate absorption depth on parsec scales \citep{roche2007}. The optical spectrum has strong [O\,{\sc iii}] and [N\,{\sc ii}] narrow lines \citep{zaw2009}. The log([O\,{\sc iii}]/H$\beta$) and log([N\,{\sc ii}]/H$\alpha$) line 
ratios, 0.88 and $-0.09$ respectively \citep{kewley2001}, place it firmly in the Seyfert region of the BPT diagram. 

NGC\,5506 is close to edge-on, 
with an inclination of 75$^{\circ}$, above and below which ionized gas is in outflow within cones with an opening angle of $\approx$\,80$^{\circ}$ \citep{maiolino1994}. These outflows are consistent with the distribution of enhanced sigma values we observe in the ionized gas in a vertical structure crossing
the galaxy plane and opening to the north and south of it (Fig.\,\ref{ngc5506}). The outflow to the north is also well traced by the [Fe\,{\sc ii}] flux distribution (its highest emission shows an approximately conical shape), while the H$_2$ flux distribution seems to better trace the gas in the galaxy plane, as it is oriented along this direction.

{As observed in Fig.\,\ref{espectros}, the  nucleus of NGC\,5506 shows a steep red continuum over the $2.1-2.4$\,$\mu$m spectral range that can be attributed to a blackbody source with temperature $\approx$\,1000\,K. Although large scale dust is observed in the central region of this galaxy, possibly due to its high inclination and as seen in the $V-H$ dust map from \citet{martini2003}, the origin of the K-band continuum is most probably hot dust emission from the AGN torus, as already observed for other active galaxies \citep[e.g.][]{riffel09-4151,burtscher15,diniz18}.


\subsection{NGC\,5899}
\label{sngc5899}
This inclined SAB(rs)c galaxy (Fig.\,\ref{ngc5899}) is reported to be in a pair and presents an optical Seyfert 2 spectrum \citep{devaucouleurs91}. The stellar kinematics derived by \citet{riffel2017} gives a PA $\approx$\,24$^{\circ}$, that is somewhat distinct from that obtained by us for the molecular gas (Paper B), of $\approx$\,4$^{\circ}$. The kinematics of the ionized gas is very different from that of the molecular gas 
(Fig.\,\ref{ngc5899}):  while the later seems to be dominated by rotation in the inclined galaxy plane, with blueshifts to the north and redshifts to the south, the ionized gas kinematics shows velocities that are opposite to that observed in the H$_2$ velocity field. A possible interpretation is that the ionized gas is tracing an outflow at PA $\approx$\,0$^{\circ}$, as supported 
also by the north-south elongation observed mostly in the [Fe\,{\sc ii}] flux map and by the enhanced velocity dispersion observed at these locations.

\subsection{Global properties}

\subsubsection{Surface mass density distributions}
Fig.\,\ref{mass} shows that the molecular gas seems to be more evenly distributed over the field of view than the ionized gas, which is more concentrated towards the nucleus and shows a more patchy and sometimes collimated mass distribution.
The more peaked distribution of the ionized gas is clearly seen in the average surface  mass density profiles shown in the third column of  Fig.\,\ref{mass} for the galaxies NGC\,788, Mrk\,607, NGC\,3227 and NGC\,5506. In the cases of NGC\,3516
and NGC\,5899, the average profiles of the molecular and ionized gas surface mass densities are similar to each other. The prevalence of more concentrated ionized gas profiles than those in H$_2$ can be attributed to the fact that the neutral gas rapidly absorbs the ionizing photons from the AGN, thus concentrating in its vicinity and/or along the ionization axis -- the preferred direction of escape of the AGN radiation.

The H$_2$, on the other hand, has another source of excitation, as shown in our previous works \citep[e.g.][]{ilha2016}, where the typical temperature is $\approx$2000\,K and the line ratios suggest thermal excitation. The heating to excite the rotational and vibrational modes of the H$_2$ molecule can be attributed to X-rays originating in the AGN. These X-rays penetrate deep along the galaxy plane in every direction heating the region and exciting the H$_2$ molecule. In addition, in the regions closest to the AGN, where the temperature can reach $\sim15.000$\,K, the H$_2$ molecules are destroyed, and we only see those more deeply embedded in the circumnuclear dust, where they are shielded from the strongest AGN radiation.


{We thus conclude that the distinct nature of the excitation of the ionized and the molecular gas and different physical conditions of the regions where they originate can explain the difference in the gas surface mass density profiles.}

\subsubsection{Gas kinematics}

{In this section we discuss briefly the global properties of the gas kinematics, with the goal of just highlighting apparent correlations among properties and differences between the molecular and ionized gas kinematics. As we already pointed out, we will leave the modeling of the velocity fields, the quantification of  possible inflows and outflows, as well as the determination of the impact of the outflows on the host galaxy (feedback) to Paper B.} 

A global property of the gas kinematics, as seen in Figs.\,\ref{ngc788}--\ref{ngc5899}, is that the warm H$_2$ velocity field is dominated by rotation in the plane of the galaxy. Although the ionized gas velocity fields show also signatures of rotation, the rotation pattern is distorted due to the presence of additional components that can be attributed to outflows and which are usually associated with an increase in the velocity dispersion.

{The above results are in agreement with those of previous studies by our AGNIFS group, in which we have additionally reported, for H$_2$, the presence of inflows in a number of galaxies. We have also found that outflows are most easily seen in the ionized gas \citep{thaisa2019} and that this gas shows a more centrally concentrated flux distribution than that of H$_2$ \cite{riffel2018}, also in agreement with what we found in the present paper.}

{The most frequent presence of outflows in the ionized gas than in the molecular gas can be understood as due to the fact that the ionized gas is mostly concentrated along the ionization axis, where it can be easily pushed by an AGN outflow. The molecular gas, on the other hand, is usually destroyed by the hard AGN radiation at these locations. Thus, although the H$_2$ flux distributions appear also surrounding the nucleus, it is not co-spatial with the ionized gas, as it originates in gas at temperatures much lower (2000K) than those characteristic of the ionized gas (10000K). This is also supported by the lower velocity dispersion of H$_2$. Clear examples are the cases of NGC\,4151 \citep{storchi2009} and NGC\,1068 \citep{barbosa14} that show H$_2$ flux distributions surrounding the nucleus but avoiding the ionization axis. On the other hand, molecular gas outflows are observed in a number of cases, and in distant AGN hosts \citep{emonts2017}. A possibility in these cases is that a nuclear outflow can push the surrounding gas including the H$_2$ gas which may be shielded from the ionizing radiation by dust, for example. We hope to re-visit this point in Paper B.}}


Regarding the presence of outflows, we have found signatures of them in most galaxies of our sample, most clearly observed in the [Fe\,{\sc ii}] kinematics: (1) in NGC\,788, along the east-west direction; (2) in NGC\,3227, towards the northeast, as suggested by distortions in the velocity field and patterns in $h_3$ and $h_4$; (3) in NGC\,3516, from the south-east to the north-west, as suggested by the disturbed velocity field and enhanced velocity dispersion; (4) in NGC\,5506, along north-south, as suggested by the enhanced velocity dispersions; (5) and in NGC\,5899, towards north and south, as indicated by the opposite velocity field between the ionized and molecular gas.


Another global property of the sample is the fact that the $h_3$ Gauss-Hermite moment shows an inverse correlation with the V$_{\rm LOS}$ in a number of cases: (1) in the Pa$\beta$ line for NGC\,788; (2) in Pa$\beta$ and [Fe\,II] for Mrk\,607; (3) in Pa$\beta$ and H$_2$ in NGC\,3227; (4) In Pa$\beta$ and H$_2$ for NGC\,3516; (5) in Br$\gamma$ and H$_2$ in NGC\,5506; (6) in H$_2$ in NGC\,5899.  This means that there are red wings in centrally blueshifted profiles and blue wings in redshifted ones. One possible explanation is the effect known as ``asymmetric drift" \citep{westfall2007}: gas rotating in the galaxy plane gives origin to the ``main"  velocity field -- corresponding to the emission-line peaks, while tenuous gas at higher latitudes rotating with lower velocity -- thus lagging behind the rotation in the plane -- gives origin to the wings of the profiles.  

{We also observe in some cases, an inverse correlation between the $h_4$ and the $\sigma$ values -- mostly positive $h_4$ values at the locations with low $\sigma$, and only a few negative values at locations with high $\sigma$. This has been observed in: (1) H$_2$ for NGC\,788; (2) in the three emission lines for NGC\,3227; (3) in Pa$\beta$ and H$_2$ for NGC\,3516; (4) in Br$\gamma$ for NGC\,5506; (5) in the three emission lines for NGC\,5899. 

Positive $h_4$ values indicate profiles more ``peaky" than a Gaussian curve, but with broader wings, while negative values indicate profiles less peaky and with less extended wings than a Gaussian. We tentatively interpret the positive $h_4$ values as being mostly due to the emission of gas rotating in the disk, with low $\sigma$ values, while the wings could originate in diffuse gas emission, that may extend to high latitudes, increasing the range of velocities probed by the emission. The high $\sigma$ and negative $h_4$ values are rarer, and could arise in spatially unresolved double components, and may be related to the presence of an outflow, as seems to be the case for the [Fe\,II] line in NGC\,3227, for example (Fig.\,\ref{ngc3227}).}

\subsubsection{Total gas masses and implied star formation rates}

In order to evaluate if the gas reservoirs accumulated in the inner few hundred pc of these galaxies have enough mass to feed the AGN, we first estimate the AGN accretion rates $\dot{m}$ using:

\begin{equation}
\dot{m}=\frac{L_{\rm bol}}{c^2\eta},
\end{equation}

\noindent where $L_{\rm bol}$ is the bolometric luminosity of the AGN, $c$ is the light speed, and $\eta$ is the conversion efficiency of rest mass of the accreted material into radiation, that we have adopted as $\eta$=0.1. {The bolometric luminosities were determined by \citet{riffel2018} based on the hard X-ray luminosities. The only exception is the case of Mrk\,607, for which the [O\,{\sc iii}] luminosity was used instead, adopting a bolometric correction factor of 3500 \citep{heckman2004}. The resulting accretion rates are presented in Table\, \ref{tabfinal}.}

%


\begin{table}
\centering

\vspace{0.3cm}
\caption{Mass accretion rates $\dot{m}$, average SFR surface densities $\Sigma_{\rm SFR}$, and total SFRs.}
\begin{tabular}{c c c c}
\hline
Galaxy & $\dot{m}$ & $<\Sigma_{\rm SFR}>$ & SFR \\
	& 10$^{-3}$\,M$_{\odot}$\,yr$^{-1}$ & 10$^{-3}$\,M$_{\odot}$\,yr$^{-1}$kpc$^{-2}$ & 10$^{-3}$ M$_{\odot}$\,yr$^{-1}$ 	\\
\hline
NGC788  & 49.2  & 2.5$\pm$1.2  &   1.4$\pm$0.7 \\
Mrk607  & 0.1   & 26.5$\pm$29.2  &   0.6$\pm$0.7  \\
NGC3227 & 4.1   & 20.1$\pm$2.5  &   0.9$\pm$0.2 \\
NGC3516 & 27.9  & 1.1$\pm$0.8   &   0.1$\pm$0.1 \\
NGC5506 & 37.0   & 260$\pm$18.0   &   31.2$\pm$2.2 \\
NGC5899 & 2.4   & 1.0$\pm$1.0    &   0.2$\pm$0.2 \\
\hline
\end{tabular}
\label{tabfinal}
\end{table}
 
As the estimated mass accretion rates to the AGN are in the range 0.1\,--\,50\,$\times\,10^{-3}$\,M$_{\odot}$\,yr$^{-1}$, considering an AGN activity cycle of 10$^7$ -- 10$^8$ yr, and assuming that most of the ionized and molecular gas are concentrated within the inner few 100 pc of the galaxies (Fig.\,\ref{mass}), it can be concluded that the ionized gas mass alone would be enough to feed the AGN. Nevertheless, Table\,\ref{tabmass} shows that the estimated masses of the cold molecular gas are larger, and range from 10$^7$ to 10$^8$ M$_{\odot}$; therefore, there seems to be at least $\approx$10$^2$ times more gaseous mass in the inner few 100\,pc of these galaxies than that necessary to feed the AGN. 

 {The fate of the gas that is not used to feed the AGN can be: (1) be consumed by star formation; (2) be pushed away by AGN feedback; (3) be pushed away by stellar feedback \citep[e.g.][]{hopkins2016}. Most probably more than one process will occur. 
 
  Here we will discuss only the possibility that the gas accumulated in the nuclear region will lead to the formation of new stars, thus calculating the star formation rate (SFR) in the inner few 100 pc. The  feedback due to the observed outflows and possibility of stellar feedback will be discussed in Paper B.} 
 
 \citet{schmidt1959} has shown that the SFR is directly related to the gas density, and later \citet{kennicutt1998} proposed a relation between the SFR surface density $\Sigma_{\rm SFR}$ and the ionised gas mass surface density $\Sigma_{\rm H{II}}$, as follows:

\begin{equation}
\frac{\Sigma_{\rm SFR}}{\rm M_{\odot}yr^{-1}kpc^{-2}}=(2.5\pm 0.7) \times 10^{-4} \left(\frac{\Sigma_{\rm HII}}{\rm M_{\odot}pc^{-2}} \right)^{1.4}, 
\end{equation}

Using the relation above, we obtain the mean values for $\Sigma_{\rm SFR}$ listed in Table\,\ref{tabfinal}, which range from 1$\times$10$^{-3}$ to 0.26\,M$_{\odot}$ yr$^{-1}$ kpc$^{-2}$. Using the areas quoted in Table\,\ref{tabmass} for the regions occupied by the ionized gas we estimate total star formation rates for the area covered by our observations (radius of $\approx$\,300 pc)  in the range 10$^{-4}$ -- 10$^{-2}$ M$_{\odot}$ yr$^{-1}$ (Table\,\ref{tabmass}). These values are within the range of values observed for the inner few 100 pc of galaxies and circumnuclear star-forming regions \citep{shi2006,dors2008,falcon2014,tsai15,riffel2016sfr}. 

{We therefore conclude that the mass reservoirs in the inner 300\,pc of the sample galaxies can not only power the central AGN but also form new stars at low SFR ($\le\,$10$^{-2}$ M$_{\odot}$ yr$^{-1}$). The presence of recently formed stars in the inner few 100 pc of AGN is supported by the observation of low-stellar velocity dispersion ($\sigma_*$) structures in 10 of 16 galaxies of our sample for which we could measure $\sigma_*$ \citep{riffel2017}.}

\section{Conclusions}
\label{sec:conclusions}
We have mapped the ionized and molecular gas flux distributions, excitation and kinematics in the inner kpc of 6 nearby active galaxies using adaptative optics assisted NIR J- and K-band integral field spectroscopy obtained with the Gemini NIFS instrument. The main conclusions of this work are listed below.

\begin{itemize}
\item The flux distributions are usually distinct for the ionized and molecular gas:  while the former is more concentrated and sometimes collimated along a preferred axis, the latter is distributed more uniformly over the galaxy plane. These flux distributions lead to azimuthally averaged surface mass density profiles steeper for the ionized gas than for the molecular gas. We attribute this difference to the different excitation mechanisms: while the ionized gas is excited by the AGN radiation in regions with temperatures of about 10000\,K, close to the AGN, the molecular gas is thermally excited in regions of lower temperatures of about 2000\,K, that extend farther from the nucleus;

\item The gas kinematics is also distinct: while the molecular gas is mostly rotating in the galaxy plane with low velocity dispersions, the ionized gas frequently shows other components associated with higher velocity dispersions and distorted velocity fields suggesting outflows; 

{\item Signatures of outflows are mostly observed in the [Fe\,{\sc ii}] kinematics of NGC\,788,  NGC\,3227, NGC\,3516,  NGC\,5506, NGC\,5899. The modelling of the gas kinematics and quantification of the mass outflow rates and powers will be presented in a forthcoming paper (Paper B);}

\item There is usually an inverse correlation between the $h_3$ Gauss-Hermite moment and the velocity field: positive values (red wings) are associated to blueshifts and negative values (blue wings) are associated to redshifts. This can be understood as due to the gas rotating closer to the galaxy plane originating the ``main" velocity field -- corresponding to the emission-line peaks, while tenuous gas at higher latitudes rotating with lower velocities originate the ``lagging" wings; 

\item There is in some cases also an inverse correlation between $h_4$ and the velocity dispersion maps: low values of $\sigma$ correspond to positive values of $h_4$ (peaky profiles with extended wings), while high $\sigma$ values correspond to negative values of $h_4$ (boxy profiles). We see mostly positive $h_4$ values, which we attribute to gas rotating in the disk (low $\sigma$), on which less luminous emission of hotter gas at higher latitudes gives origin to the broad wings;

\item Although the excitation will be further analised in Paper B, general trends observed in the emission-line ratios are: (1) an increase in [Fe\,{\sc ii}]$\lambda$1.2570$\mu$m/Pa$\beta$ in association with higher velocity dispersion indicating contribution from shocks, and an increase in H$_2\,\lambda\,2.12\mu$m/Br$\gamma$ outwards, attributed to destruction of the H$_2$ molecule close to the AGN due to its strong radiation;


\item The integrated mass of ionized gas within the inner $\approx$\,300\,pc radius ranges from 1.8$\times10^5$\,M$_{\odot}$ to 1.9$\times10^7$\,M$_{\odot}$, while that of warm molecular gas is $\sim$\,10$^{3-4}$ times lower, and the estimated mass in cold molecular gas is $\sim$\,10$^2$ times higher;


\item The {\color{black}average} ionized gas surface mass density ranges from 2.8\,M$_{\odot}$\,pc$^{-2}$ to 140\,M$_{\odot}$\,pc$^{-2}$, while for the warm molecular gas it is $\sim$\,10$^{3-4}$ times lower, and that estimated for the cold molecular gas is $\sim$\,10$^2$ times higher.

 \item The AGN accretion rates in our sample range from 0.1$\times$10$^{-3}$M$_{\odot}$ yr$^{-1}$ to 49$\times$10$^{-3}$M$_{\odot}$ yr$^{-1}$; considering an activity cycle of duration 10$^7$ -- 10$^8$\,yr, it can be concluded that there are $\approx$\,10$^2$ times more gas in the inner few hundred pc of the galaxies than needed to feed the AGN over a duty cycle;

\item If most of this gas will lead to star formation in the inner few 100 pc, we estimate  star formation rates in the range 1--260$\times$ 10$^{-3}$ M$_{\odot}$ yr$^{-1}$ which is within the range of typical values observed for circumnuclear star-forming regions in nearby galaxies;  

\item {The mass reservoirs in the inner few 100 pc of these galaxies are thus enough to power both the central AGN and star formation. But our observations show also that at least part of this gas is being pushed away by an AGN-driven outflow or supernovae winds, which we will further investigate via the analysis of the gas kinematics in Paper B.}


\end{itemize}

 This paper is the thirteenth of a series by our group AGNIFS in which we have been mapping in detail AGN feeding and feedback processes in nearby galaxies using adaptative optics assisted integral field observations with  NIFS. And it is the third in which we aim to characterize a global sample of 20 nearby active galaxies, the first one in which we characterized the stellar kinematics \citep{riffel2017} and the second in which we presented the sample and mass density profiles \citep{riffel2018}. The observations are scheduled to be concluded in 2019. Further and more detailed analysis of the the gas kinematics and  excitation for the six galaxies presented in this work will be presented in the forthcoming Paper B.

\section*{Acknowledgments}
We thank an anonymous referee for valuable suggestions which helped to improve the paper. 
This study was financed in part by the Coordena\c c\~ao de
Aperfei\c coamento de Pessoal de N\'ivel Superior - Brasil (CAPES) -
Finance Code 001, Conselho Nacional de Desenvolvimento Cient\'ifico e Tecnol\'ogico (CNPq) and Funda\c c\~ao de Amparo \`a pesquisa do Estado do RS (FAPERGS).
This work is based on observations obtained at the Gemini Observatory, 
which is operated by the Association of Universities for Research in Astronomy, Inc., under a cooperative agreement with the 
NSF on behalf of the Gemini partnership: the National Science Foundation (United States), the Science and Technology 
Facilities Council (United Kingdom), the National Research Council (Canada), CONICYT (Chile), the Australian Research 
Council (Australia), Minist\'erio da Ci\^encia e Tecnologia (Brazil) and south-east CYT (Argentina).  







\bibliographystyle{mnras}
\bibliography{refs} 





\bsp	
\label{lastpage}
\end{document}